\documentstyle[prd,aps,preprint]{revtex}
\tightenlines
\input epsf.tex
\def\DESepsf(#1 width #2){\epsfxsize=#2 \epsfbox{#1}}

\newcommand{\be}{\begin{equation}}
\newcommand{\ee}{\end{equation}}
\newcommand{\bea}{\begin{eqnarray}}
\newcommand{\beas}{\begin{eqnarray*}}
\newcommand{\eea}{\end{eqnarray}}
\newcommand{\eeas}{\end{eqnarray*}}
\newcommand{\ba}{\begin{array}}
\newcommand{\ea}{\end{array}}

\begin{document}

\draft
\preprint{\vbox{
\hbox{UMD-PP-00-043}}}

\title{Type II Seesaw and a Gauge Model for the Bimaximal Mixing
Explanation of Neutrino Puzzles}
\author{ R. N. Mohapatra$^1$\footnote{e-mail:rmohapat@physics.umd.edu},
A. P\'erez-Lorenzana$^{1,2}$\footnote{e-mail:aplorenz@Glue.umd.edu} 
and C. A. de S. Pires$^{1}$\footnote{e-mail:cpires@physics.umd.edu}}

\address{
$^1$ Department of
Physics, University of Maryland, College Park, MD, 20742, USA\\
$^2$  Departamento de F\'\i sica,
Centro de Investigaci\'on y de Estudios Avanzados del I.P.N.\\
Apdo. Post. 14-740, 07000, M\'exico, D.F., M\'exico. }

\date{November, 1999}

\maketitle

\begin{abstract}
{We present a gauge model for the bimaximal mixing
pattern among the neutrinos that explains both the
atmospheric and solar neutrino data via large angle vacuum oscillation
among the three known neutrinos. The model does not include righthanded
neutrinos but additional Higgs triplets which acquire naturally small
vev's due to the type II seesaw mechanism.
A combination of global $L_e-L_{\mu}-L_{\tau}$ and
$S_3$ symmetries constrain the mass matrix for both
charged leptons and neutrinos in such a way that
the bimaximal pattern emerges naturally at the tree level and needed
splittings among
neutrinos at the one loop level. This model predicts observable branching 
ratios for $\tau\rightarrow \mu \mu\mu$, which could be used to test it.}
\\[1ex]
\end{abstract}

\vskip0.5in

\section{Introduction}

    The observation of a deficit as well as of a zenith angle dependence
in the flux of atmospheric muon neutrinos 
by the Super-Kamiokande~\cite{sk,atmos}  collaboration has provided strong
evidence that there are oscillations among the known neutrino species. The
five solar neutrino experiments~\cite{expt,superK}  have added to this
sense of excitement by their long standing result that there is also a
deficit of the solar neutrinos, which can be given a simple explanation
in terms of neutrino oscillations~\cite{rev}. It thus appears certain
that neutrinos have mass and they mix among each other. Although the
details are fuzzy on the exact nature of the oscillations needed
for the purpose,
several very interesting scenarios exist. In particular, if the only
laboratory indication of the neutrino oscillation by the Los Alamos
collaboration (LSND)~\cite{LSND} is not included in the picture, there is
a mixing scheme known as the bimaximal mixing, where
both the solar and atmospheric data are explained by large mixing 
among the three known neutrinos~\cite{gold}. In this picture, solar
neutrino puzzle could either be solved via the large angle MSW
mechanism~\cite{MSW} or via the vacuum oscillation
mechanism~\cite{bahcall} depending on the mass difference between the
muon and the electron neutrinos. In this paper we will assume the
vacuum oscillation between the $\nu_e$ and
$\nu_{\mu}$ as the solution, which requires that their mass difference
square must be
$\sim 10^{-10}$ eV$^2$.  The observed electron energy distribution as
well as some hints of bi-annual variation of the solar neutrino flux by
the Super-Kamiokande collaboration may be pointing in this direction.

If we accept this particular resolution of the neutrino puzzles, 
two major theoretical challenges emerge: one, how does one get naturally
a theory that leads to the bimaximal mixing matrix and two, how does
the same framework explain the tiny mass difference square ($\sim
10^{-10}$ eV$^2$) needed for the purpose without
fine tuning of parameters ? Our goal in this paper is to provide a simple
model that generates both these features of the neutrino physics.
Note that having a neutrino mass matrix of the right form to generate the
bimaximal mixing pattern is by itself not sufficient since the desired
mixing matrix is a combination of both the neutrino and the charged lepton
mixing matrices i.e., $U^{\dagger}_{\ell}U_{\nu}$. Often it is {\it 
assumed} that $U_{\ell}= {\bf 1}$ by appropriately choosing couplings in
a theory to have certain values. This is of course not technically  
natural since radiative corrections could induce arbitrary values for
those parameters thereby upsetting the neutrino mixing pattern. So what is
really needed is (i) a neutrino Majorana mass matrix of the right form to
generate the bimaximal mixing and (ii) a diagonal charged lepton mass
matrix . It is the goal of this paper to present a model that satisfies
both these criteria naturally. In this respect our model is different from
others discussed in the literature (see later for detailed comparison).

One of the key ingredient in our work is the type II seesaw mechanism
where the vev of
a triplet Higgs becomes ultrasmall due to the presence of a high scale in
the theory\cite{moh1}. The presence of additional global symmetries in the
model lead to a pattern of neutrino masses that leads to the bimaximal
mixing among neutrinos while keeping flavor mixing among the charged
leptons to be zero so that the bimaximal pattern dictated by the neutrino
mass matrix that emerges is indeed natural.

Using the definition of the mixing matrix as 
 \begin{eqnarray}
 \left(\begin{array}{c} \nu_e\\ \nu_{\mu} \\ \nu_{\tau}\end{array}
 \right)=~ U_{\nu}\left(\begin{array}{c} \nu_1 \\ \nu_2 \\ \nu_3
 \end{array} \right) ;
 \end{eqnarray}
the bimaximal mixing corresponds to the mixing matrix $U_{\nu}$ given
by\cite{gold}
 \begin{eqnarray}
 U_{\nu}=\left(\begin{array}{ccc}
 \frac{1}{\sqrt{2}} & -\frac{1}{\sqrt{2}} & 0\\
 \frac{1}{2} &\frac{1}{2} &\frac{1}{\sqrt{2}} \\
 \frac{1}{2} &\frac{1}{2} &-\frac{1}{\sqrt{2}}
 \end{array}\right) .
 \end{eqnarray}
As far as the neutrino masses go, they may be fully or partially 
degenerate or hierarchical as long as the mass differences fit the
desired values. As mentioned, a 
convincing theoretical explanation of this elegant mixing pattern seems so
far to have been elusive, although there exist many interesting
attempts\cite{theory,ma,jr,barb,more}. The problem becomes even more
challenging when we 
demand that the solar neutrino puzzle be solved by the vacuum oscillation 
mechanism.

In this letter, we use the type II seesaw mechanism in conjunction with
the global symmetry $S_3\times U(1)_{e-\mu-\tau}$ to show that both these
properties can be realized in a natural manner. This leads us to a
neutrino mass pattern where $m_{\nu_{\tau}}\ll m_{\nu_{\mu}}\simeq
m_{\nu_e}\simeq \sqrt{\Delta m^2_{ATMOS}}\simeq 0.05$ eV and a
generalized bimaximal pattern given by:
 \begin{eqnarray}
 U_{\nu}=\left(\begin{array}{crc}
 \frac{1}{\sqrt{2}} & -\frac{1}{\sqrt{2}} & 0\\
 \frac{\cos\theta}{\sqrt{2}} &\frac{\cos\theta}{\sqrt{2}}
 &-\sin\theta \\
 \frac{\sin\theta}{\sqrt{2}} &\frac{\sin\theta}{\sqrt{2}}
 &\cos\theta \end{array}\right) .
 \label{Uv}
 \end{eqnarray}

\section{The Model}
 We consider an extension of the standard model where the fermion content
is left unaltered but with a Higgs sector extended as follows: three
doublets $\phi_0, \phi_1, \phi_2$, two triplets with $Y=2$ denoted by
$\Delta_{1,2}$ and a charged isosinglet with $Y=+2$. The model has an
$S_3$ symmetry (i.e. permutation group on three elements), under which the
particles are assigned as shown in Table I.
 \begin{center}
 \begin{tabular}{|c||c|}\hline
 Fields & $S_3$ transformation \\ \hline
 ($L_{\mu}, L_{\tau}$) & 2 \\
 ($\mu_R, \tau_R$) & 2\\
 $L_e, e_R$ & 1 \\
 ($\phi_1, \phi_2$) & 2 \\
 $\phi_0$ & 1\\
 ($\Delta_1, \Delta_2$) & 2 \\
 $\eta^+ $ & 1' \\ \hline
 \end{tabular}
{ \quote \small 
{\bf Table I}: Transformation properties of the fields in the model under
the $S_3$ group.}
 \end{center}

We also impose an $L_e-L_{\mu}-L_{\tau}$ symmetry on the model. The Yukawa
part of the Lagrangian in the leptonic sector consistent with these
symmetries can be written as:
 \begin{eqnarray}
 {\cal L}_Y &=& 
 h_1 \left(\bar{L}_{\mu}\mu_R + \bar{L}_{\tau}\tau_R\right) \phi_0 
 +  h_2 \left[(\bar{L}_{\mu}\mu_R - \bar{L}_{\tau}\tau_R)\phi_1  
 + (\bar{L}_{\mu}\tau_R + \bar{L}_{\tau}\mu_R)\phi_2\right] \nonumber \\
 & &+ h_e \bar{L}_e e_R \phi_0 
 + f L_e \left( L_{\mu}\Delta_1 + L_{\tau}\Delta_2 \right) 
 + f' L_{\mu}L_{\tau}\eta + h. c.
 \label{llag}
 \end{eqnarray}

We will show later by a detailed examination of the Higgs potential for
the system that there is a domain of parameters for which we get the
following vevs of the fields:
\begin{equation}
\langle\Delta^0_{1,2}\rangle ~=~v^T_{1,2} ;\quad
\langle\phi_0\rangle = v_0; \quad 
\langle\phi^0_1\rangle = v_1; 
\quad \langle\phi^0_2\rangle=0 .
\end{equation}
Clearly, the pattern of $\phi$ vevs leads to a diagonal mass matrix for
the charged leptons whereas the $\Delta$ vev's leads to a Majorana mass
for the neutrinos of the form:
\begin{eqnarray}
M_{\nu}=\left(\begin{array}{ccc}
0 & m_1 & m_2 \\
m_1 & 0 & 0 \\
m_2 & 0 & 0 \end{array}\right)  .
\end{eqnarray}
As a consequence, diagonalization of the neutrino mass matrix is solely
responsible for the neutrino mixings and one obtains the pattern given in
the generalized bimaximal form~\cite{barb,more} (see Eq. (\ref{Uv})), with
$\tan\theta = m_2/m_1 = v_2^T/v_1^T$.

Let us now address the question of neutrino masses. Clearly, to understand
the small neutrino masses, one must have a tiny value for the vev of the
$\Delta$ fields. This is achieved by the type II seesaw
mechanism~\cite{moh1}. This is a generic mechanism, which can be
illustrated by the following simple model that has only one $\phi$ and one
$\Delta$. Consider the following Higgs potential for this
system~\cite{moh1}:
\begin{eqnarray}
V(\phi, \Delta) &=& M^2 \Delta^{\dagger}\Delta -\mu^2\phi^{\dagger} \phi 
+ \lambda_{\phi} (\phi^{\dagger}\phi)^2   
+ \lambda_{\Delta}\Delta^{\dagger}\Delta  \nonumber \\
&&+\lambda_{\phi\Delta}\Delta^{\dagger}\Delta\phi^{\dagger}\phi +
M_{\Delta\phi\phi}\Delta^{\dagger}\phi\phi ~ + ~ h.c.
\end{eqnarray}
Let us choose $\mu\sim 100$ GeV and $M\sim M_{\Delta\phi\phi}\gg \mu$; in
this case, the vev of $\langle\phi^0\rangle\approx \mu$ whereas the vev of 
$\langle\Delta^0\rangle \sim \frac{\mu^2}{M}\ll \mu$.
This mechanism has been labeled type II seesaw and we see that
if $M\simeq 10^{14}$ GeV, then we get $\langle\Delta^0\rangle\simeq 0.14$ eV. 
In the presence of more $\Delta$ fields and extra symmetries that our model 
has this mechanism still operates and we have a small mass (in the $0.1$ eV
range) for the $\nu_e$ and $\nu_{\mu}$. The third eigenstate has zero
mass. The $\nu_{\mu}$ and $\nu_{\tau}$
get mixed in the tree level and the mixing angle is near maximal unless we
do fine tuning. As far as the $\nu_{\mu}$ and $\nu_e$ are concerned, they
are degenerate and have opposite CP; therefore at the tree level their
mass difference
vanishes. We will show below that their mass difference arises at the one
loop level but due to the presence of the high mass scale that gave rise
to the type II seesaw mechanism, the mass difference $\Delta m^2_{e-\mu}$
is naturally suppressed to the level of $10^{-10}$ eV$^2$ without any 
unnatural fine tuning. The tree level mass matrix already explains the
atmospheric neutrino puzzle due to the type II seesaw. 

Let us now turn to the explanation of the one loop contribution to the
neutrino mass matrix. This is where the role of the $\eta$ boson becomes
important. Let us first note that the masses of the doublet bosons are of
order of the electroweak scale ($~100$ GeV) since their vevs must be of
that order whereas that of the singlet $\eta$ and the
 triplet bosons are heavy (i.e., of order $10^{14}$ GeV). In order to
generate the mass difference between the $\nu_e$
and $\nu_{\mu}$, we need nonzero entries for the $\mu\mu$ or $ee$ element.
Both of them will violate the $L_e-L_{\mu}-L_{\tau}$ symmetry. This
breaking is introduced by a soft term in the potential $\eta^* 
\phi_1\phi_2$ since $\eta$ has $L_e-L_{\mu}-L_{\tau}= -2$ and $\phi$'s are
neutral under this global symmetry. This is a dimensional coupling and in
accordance with our principle above that all fields which are not
involved in the process of electroweak symmetry breaking are superheavy,
we will choose this to be of order $M$. This
leads to  one loop graphs as in Figs. 1 and 2, 
which produce a neutrino mass
matrix as follows:
\begin{eqnarray}  
M_{\nu}=\left(\begin{array}{ccc}
0 & m_1 & m_2 \\
m_1 & m_{\mu\mu} & m_{\mu\tau} \\
m_2 &  m_{\mu\tau} & m_{\tau\tau} \end{array}\right) ;
\end{eqnarray}
where $m_{\mu\tau}\simeq m_{\mu\mu}\sin\theta\cos\theta$ and
\begin{eqnarray}
 m_{\mu\mu}\simeq
\frac{f'h_2}{16\pi^2}\frac{m_{\tau}M_{\eta\phi\phi}v}{M^2} .
\end{eqnarray}
For $M\simeq M_{\eta\phi\phi}$ and $f'h_2\simeq 10^{-4}$, we get
$m_{\mu\mu}\sim 10^{-9}$ eV, leading to $\Delta m^2_{e\mu}\simeq 10^{-10}$
eV$^2$, as is required to explain the solar neutrino puzzle via vacuum
oscillation. Note that the tau neutrino picks up very tiny mass at 
one loop ($\sim 10^{-10}$ eV). This completes the derivation of the main
result of our paper. The important point to note is that we need to choose
the Yukawa couplings $f'$ and $h_2$ individually only of order $10^{-2}$.

Let us now compare our model with two existing ones\cite{ma,jr} in the
literature. The model of Ma\cite{ma} has a similar
field content to ours in all respects except that there is only one
triplet field as against two in our paper. Thus inspite of the $S_3$
symmetry in that paper, the
tree level mass matrix is very different and one does not have the
bimaximal pattern at the tree level from the neutrino sector sector alone
as in our case. On the other hand, the work of Ref.\cite{jr} uses only the 
$L_e-L_{\mu}-L_{\tau}$ symmetry, which allows arbitrary mixing angles in
the charged lepton sector, which have to be set to zero at tree level.
The addition of the $S_3$ symmetry as we do in this paper helps us to keep
the charged leptons diagonal. There are also other major differences
between the work of \cite{jr} and this work in the way the detailed
dominant entries of the neutrino mass matrix arise- in our case at the
tree level
where as in \cite{jr} at the one loop model via the Zee-mechanism.

\section{Rare Tau decays}

In this section we present a test of the
model that involves flavor changing decays of the tau lepton. There are
three sources of flavor
changing effects in this model, i.e., via the exchanges of $\eta$,
$\Delta$'s and $\phi_2$. At tree level, $\eta^+$ and
$\Delta_{1,2}^{0,+}$ exchanges lead to
$\mu$ or $\tau$ decay processes that include neutrinos as final products.
However, as the masses of the exchange fields are
very heavy ($\sim M$) such processes are highly suppressed. Therefore, 
only the diagrams that involve the exchange of the doublet fields are
important.

The general $\phi$ exchange tree level diagrams for Flavor Changing
Neutral Currents (FCNC) are depicted in Figures 3 and 4. 
The internal lines in those figures
represent the contributions of the 
general mixings in the scalar sector
that come from the trilinear and quartic scalar couplings. 
In both of the figures, since one scalar 
field is necessarily a $\phi_2$ field, the $S_3$ symmetry restricts
the involved  vevs. In  Fig. 3, for instance,  it allows only 
$\langle \Delta_2\rangle$, regardless of whether the other field is 
$\phi_0$ or $\phi_1$. Thus,  we estimate the amplitude of this
diagram to be 
 \be 
 h h_2{\langle \Delta_2 \rangle M_{\Delta\phi\phi} \over 
 m_{\phi_2}^2 m_{\phi_{0,1}}^2 } 
 \simeq h \cdot 10^{-6} ~ {\rm GeV}^{-2} ; 
 \label{tree}
 \ee
where we have assumed as before that $M_{\Delta\phi\phi}\simeq M$,
$m_{\phi}\simeq v$, $M_{\Delta\phi\phi}\langle \Delta\rangle\simeq v^2 $
and taking $h_2= m_\tau/v\simeq 10^{-2}$. Here,
$h$ represent the Yukawa coupling of the  vertex on the right.

For the diagram in Fig. 4, the vevs could be
either $\langle\phi\rangle$ or $\langle\Delta\rangle$. Nevertheless, if we
choose $\langle\phi\rangle$, again
the $S_3$ symmetry play an important role by constraining 
one of them to be
$\langle\phi_2\rangle$, which  is zero, then the only contribution comes
from the $\langle\Delta\rangle$ sector which is more suppressed already
than the previous case by an extra $10^{-14}$ 
times the quartic coupling constant, which 
being dimensionless may be chosen of the order of one. There is no
compensating factor such as large $M$ as in diagram in Fig. 3. 

Thus the
only observable contribution to leptonic FCNC processes come from the
diagram in Fig. 3. Matching the external leptonic legs in Fig. 3 with the
terms in the Lagrangian, we see that only observable processes are of type
$\tau\rightarrow \mu\mu\mu$ and $\tau\rightarrow\mu e e$, which have an
amplitude of order $h 10^{-6}~{\rm GeV}^{-2}$. Summing up the two
contributions and assuming $\langle \phi_0\rangle \simeq \langle
\phi_1\rangle $ and all dimensional couplings to be same, we get $h\simeq 
\frac{g_2 m_{\mu}}{\sqrt{2}M_{W}}\simeq 10^{-3}$. This leads to the
branching ratio for the decay mode $B(\tau\rightarrow \mu\mu\mu)\simeq 
10^{-7}$. Since in our estimate we have assumed several couplings to be of
order one, the prediction is uncertain within an order of magnitude but we
do not expect it to be much smaller.
 This may be compared with the present experimental
bounds~\cite{pdg} for the decays 
$\tau\rightarrow \mu \mu\mu$ which is $10^{-6}$.
The branching ratio for the other allowed decay mode in our model
$B(\tau\rightarrow\mu e e)\simeq 10^{-12}$ is small due to the
fact that it involves the electron Yukawa coupling which is $\sim
m_e/m_W$. Therefore, the three $\mu$ rare mode is the only observable
FCNC processes in $\tau$ decays in this model and 
could therefore be used as a test.

Let us stress that, 
an interesting feature of the present model is that the
really rare process $\mu\rightarrow eee$ is automatically suppressed as it
does not appear at the tree level.  
This is because, the
only tree level coupling  among electron and muon (or tau) involves 
$\Delta^{++}$, which does not mix with  $\phi_0$,  as would be required
to get three electrons in the final states. 

At one loop order the most interesting process again appears in $\tau$
physics, i.e. the rare decay $\tau\rightarrow \mu\gamma$. The coresponding
diagram is showed in Fig. 5. Now the decay width for this process is
roughly estimated to be
 \be 
 \Gamma \simeq 
 {h_2^2 h_1^2 e^2\over 16 \pi^2}
 {\langle \Delta_2 \rangle^2 M_{\Delta\phi\phi}^2 m_\tau^5\over 
 m_{\phi_2}^4 m_{\phi_{0,1}}^4}\simeq 10^{-21}~ {\rm GeV} ;
 \ee
giving a branching ratio of about 
$\sim 10^{-9}$, which is again below the 
current experimental bound~\cite{cleo} of $10^{-6}$.

\section{Analysis of the Higgs potential }

Lets us now show how 
$\langle\phi^0_2\rangle=0$ arises naturally from the
potential. Given the irreducible representations (irreps) of $S_3$:
$2_x=(x_1,x_2)$ and $2_y=(y_1,y_2)$, we may build the following singlets
$1_{xy} = x_1 y_1 + x_2 y_2$; $1'_{xy}= x_1 y_2 - x_2 y_1$; and 
the new doublet $2_{xy} = (x_1 y_1 - x_2 y_2, x_1 y_2 + x_2 y_1)$. Using
this simple rules it is straightforward to find all possible $S_3$
and gauge invariant terms    
that include the scalar fields of the model. 
Such potential may be decomposed as 
$V= V(\phi) + V(\Delta) + V(\phi,\Delta) + V(\phi,\Delta,\eta)$.
The last term involves all the expressions containing $\eta$. They do not
contribute to the minimization of the potential, thus, it is not necessary
to show them explicitly. 

As we already discussed above, the type II seesaw formula arises from
$V(\Delta) + V(\phi,\Delta)$  by assuming large trilinear couplings. 
In this case $\langle\Delta\rangle$ becomes much much smaller than
$\langle\phi\rangle$ and then we may neglect those terms in the analysis
of the $\phi$ vevs. Defining $\Phi =(\phi_1,\phi_2)$,
the relevant terms of the potential are then 
represented by
 \bea
 V(\phi) &=& \mu^2 ~ 1_{\Phi^2} + \mu_0^2 ~ 1_{\phi_0^2} + 
 \lambda_1 ~ (1_{\Phi^2})^2 + \lambda_2 ~ (1_{\phi_0^2})^2 + \nonumber \\
  && \lambda_3 ~ 1_{\Phi^2} ~ 1_{\phi_0^2} + \lambda_3 ~ 1_{\Phi^4} + 
 \lambda_5 (1'_{\Phi^2})^2 + \lambda_6 ~ 1_{\Phi^3} ~ 1_{\phi_0};
 \eea
where $1_{x^n}$ means the singlet built by using $n$ $x$ irreps. By
examing $V(\phi)$ we may see that all the terms except the last one 
obey an accidental $U(1)_\alpha$ 
symmetry, which  makes itself evident if
we parametrize the vevs as 
 \be 
 \left(\ba {c} \langle\phi_1\rangle \\ \langle\phi_2\rangle \ea \right)
 = v
 \left(\ba {c} \cos\alpha \\ \sin\alpha \ea \right).
 \ee
Then, in  terms  of these parameters, the potential reduces to
 \be
 V(\phi) = V(v,\langle\phi_0\rangle) + 
 \lambda_6 \langle\phi_0\rangle v^3 \cos\alpha .
 \ee
If the last term was absent, the minimum would exhibit a flat direction
and an ensuing Goldstone boson. However,
 the last term breaks this extra symmetry explicitly and removes the
flatness. Moreover, from
this expression it is now straightforward to see that the potential gets
its minimum for $\alpha = 0$ ($\pi$) if $\lambda_6<0$ ($>0$), which means
that $\langle\phi_2\rangle = 0$, as we expected.
It is worth pointing out that 
this special effect in the potential does not appear on the
$\Delta$ sector, since it is just a consequence of the presence of the 
extra singlet $\phi_0$. As a matter of fact, $V(\Delta)$ is totally 
$U(1)_\theta$ symmetric while
$V(\phi,\Delta)$ contains several terms that break explicitly
such symmetry in a less trivial way, 
then avoiding a  null value of $\langle\Delta_2\rangle$, and 
giving the pattern of neutrino masses.

\section{Conclusions}

We have presented a model of neutrino masses
based on the permutation symmetry $S_3$ in conjunction with the
$L_e-L_{\mu}-L_{\tau}$ symmetry which leads to the bimaximal mixing
pattern at the tree level and explain atmospheric oscillations through
the hierarchy $m_{\nu_{\tau}}\ll m_{\nu_{\mu}}\simeq m_{\nu_e}$. The
naturalness of the bimaximal mixing follows from the fact that
the tree level mass matrix of the charged leptons is diagonal while
the neutrino Majorana mass has a specific form dictated by the type II
seesaw mechanism and the above symmetries. The soft breaking of the
$L_e-L_{\mu}-L_{\tau}$ symmetry by the scalar potential through a
coupling of the scalar doublets with an $S_3$ odd charged scalar, $\eta$, 
leads to the small entries in the neutrino mass matrix via
radiative corrections. They are responsible for the 
small splitting in $\Delta m_{e\mu}^2$ needed to
explain the solar neutrino deficit via vacuum oscillations.
This model can be tested via the rare decay $\tau\rightarrow
\mu\mu\mu$ whose branching ratio is predicted to be not too far below the
current experimental limit\cite{cleo}.

\vskip1em

{\it Acknowledgements.}   The work of RNM is supported by a grant from the
National Science Foundation under grant number PHY-9802551. The work of
APL is supported in part by CONACyT (M\'exico). The work of CP is supported
by Funda\c c\~ao de Amparo \`a Pesquisa do Estado de S\~ao Paulo (FAPESP).


\begin{figure}
\centerline{
\epsfxsize=200pt
\epsfbox{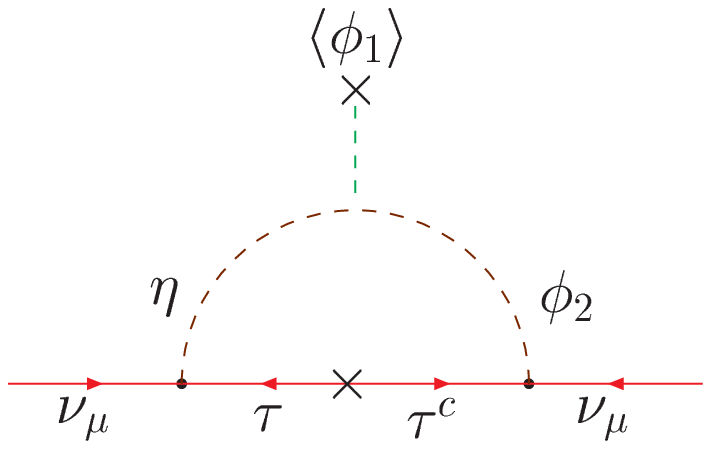}
}
\vskip1ex

\caption{One loop correction that generates the diagonal mass term 
$m_{\mu\mu}$. A similar diagram provides $m_{\tau\tau}$.}
\end{figure}
\vskip2em

\begin{figure}
\centerline{
\epsfxsize=200pt
\epsfbox{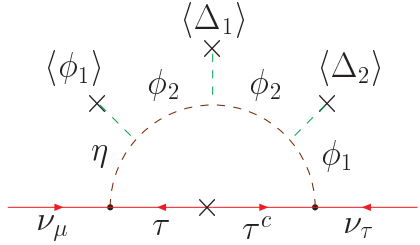}
}
\vskip1ex

\caption{One loop correction that generates the mass term 
$m_{\mu\tau}$.}
\end{figure}
\vskip2em

\begin{figure}
\centerline{
\epsfxsize=200pt
\epsfbox{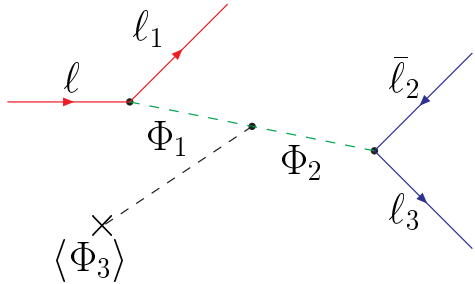}
}
\vskip1ex

\caption{Generical Feynman diagram responsible for FCC in the model at
tree level. The external lines are leptonic fields. The internal
lines could in principle be any one of the scalars involved 
in the trilinear  couplings allowed by the $S_3$ symmetry.} 
\end{figure}


\begin{figure}
\centerline{
\epsfxsize=200pt
\epsfbox{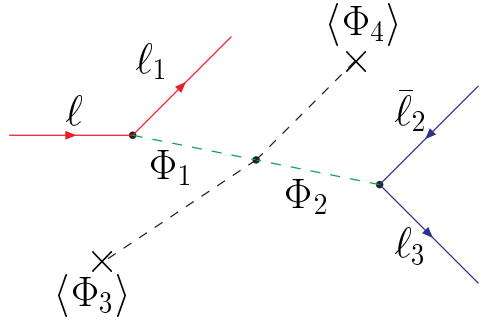}
}
\vskip1ex

\caption{Tree level FCC diagram involving the mixing produced by
quartic scalar couplings.}
\end{figure}
\vskip2em

\begin{figure}
\centerline{
\epsfxsize=200pt
\epsfbox{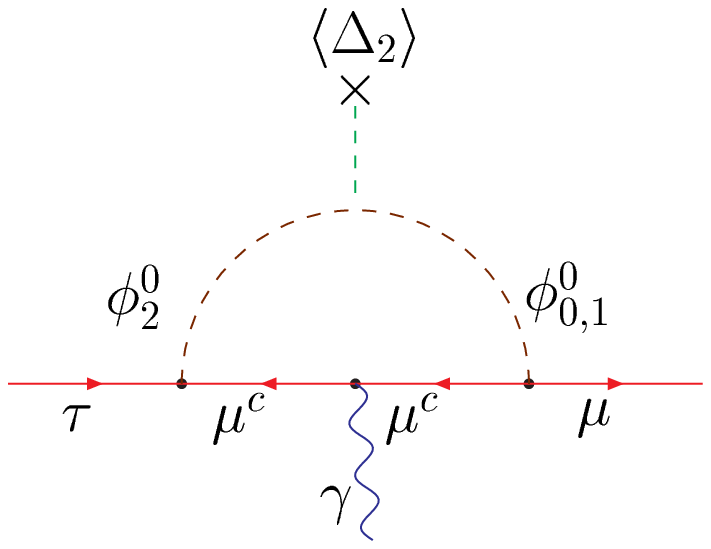}
}
\vskip1ex

\caption{One loop diagram that produce the $\tau\rightarrow \mu\gamma$
decay in the model.}
\end{figure}
\vskip2em


\begin{thebibliography}{99}

\bibitem{sk}
Y. Fukuda et al., \prl {\bf 81}, (1998) 1562; idem. {\bf 81} (1998) 1158;
\pl {\bf B436} (1998) 33.

\bibitem{atmos} 
K.S.~Hirata et al., Phys. Lett. {\bf B280} (1992) 146; 
R.~Becker-Szendy et al., Phys. Rev. D {\bf 46} (1992) 3720; 
W. W. M. Allison et al., Phys. Lett. {\bf B 391} (1997) 491;
Y. Fukuda  et al, Phys. Lett. {\bf B 335} (1994) 237.

\bibitem{expt} 
B. T. Cleveland et al. Nucl. Phys. B (Proc. Suppl.) {\bf 38} (1995) 47;
K.S.~Hirata et al., Phys. Rev. {\bf 44} (1991) 2241;
GALLEX Collaboration, Phys. Lett. {\bf B388} (1996) 384;
J. N. Abdurashitov et al., Phys. Rev. Lett. {\bf 77} (1996) 4708.

\bibitem{superK} 
Super-Kamiokande collaboration, Phys. Rev. Lett. {\bf 81} (1998)1562.

\bibitem{rev} 
For an extensive review and references, see the proceedings
of the workshop on neutrinos, NOW98 hep-ph/9906251; 
for review of
possible theories, see G. Altarelli and F. Feruglio, hep-ph/9905536;
R. N. Mohapatra, hep-ph/9910365.

\bibitem{LSND} 
C. Athanassopoulos et al. Phys. Rev. Lett. {\bf 75} (1995) 2650; 
C. Athanassopoulos et al.  \prc {\bf  58} (1998) 2489.

\bibitem{gold} 
F. Vissani, hep-ph/9708483;
V. Barger, S. Pakvasa, T. Weiler and K. Whisnant, 
 Phys. Lett. {\bf B437} (1998) 107; 
A. Baltz, A. S. Goldhaber and M. Goldhaber, \prl {\bf 81} (1998) 5730; 
M. Jezabek and Y. Sumino, Phys.Lett. {\bf B440} (1998) 327;
G. Altarelli and F. Feruglio, Phys.Lett. {\bf B439} (1998) 112; 

\bibitem{MSW}
L. Wolfestein, \prd {\bf 17} (1978) 2369; 
S.P. Mikheyev, A. Smirnov, Yad. Fiz. {\bf 42} (1985) 1441; 
 Nuovo Cimento {\bf 9C} (1986) 17.

\bibitem{bahcall} 
For a recent summary and update on the solar neutrino
puzzle, see J. Bahcall, P. Krastev and A. Y. Smirnov, 
\prd {\bf 58} (1998) 096016;
M. C. Gonzales-Garcia et al., hep-ph/9906469.

\bibitem{moh1} 
R. N. Mohapatra and G. Senjanovi\'c, Phys. Rev. {\bf D 23}, 165 (1981); 
C. Wetterich, Nuc. Phys. {\bf B 187}, 343 (1981);
E. Ma and Sarkar, Phys. Rev. Lett. {\bf 80 }, 5716  (1998).

\bibitem{theory} 
R. N. Mohapatra and S. Nussinov, Phys. Rev. {\bf D 60}, 013002 (1999); 
S. Davidson and S. F. King, Phys.Lett. {\bf B445} (1998) 191; 
C. S. Kim and J. D. Kim, hep-ph/9908435; 
C. H. Albright and S. Barr, Phys. Lett. {\bf B461} (1999) 218; 
H. Georgi and S. L. Glashow, hep-ph/9808293; 
H. B. Benaoum and S. Nasri, \prd {\bf 60} (1999) 113003; 
M. Jezabek and Y. Sumino, Phys.Lett. {\bf B457} (1999) 139;


\bibitem{ma}
E. Ma, \prl {\bf 83} (1999) 2514; hep-ph/9909249.

\bibitem{jr}
A. Joshipura and S. Rindani, hep-ph/9811252; Phys.Lett. B464 (1999) 239.

\bibitem{barb}
 R. Barbieri, L. Hall, A. Strumia and N. Weiner, JHEP 9812 (1998) 017.

\bibitem{more}
See also:
R. Barbieri, L. J. Hall, A. Strumia, Phys.Lett. {\bf B445} (1999) 407;
Y. Grossman, Y. Nir and Y. Shadmi, JHEP 9810 (1998) 007;
R. Barbieri, hep-ph/9901241;
M. Jezabek, Y. Sumino,  Phys. Lett. {\bf B457} (1999) 139;
G. Altarelli, F. Feruglio, hep-ph/9905536;
E. Kh. Akhmedov, hep-ph/9909217;
E. Kh. Akhmedov, G. C. Branco and M. N. Rebelo, hep-ph/9911364.

\bibitem{pdg}
Particle Data Group, C. Caso et al, The Eur. Phys. J. {\bf C3} (1998) 1.

\bibitem{cleo} CLEO collaboration, A. Anastassov et al., hep-ex/9908025.

\end{thebibliography}
\end{document}